\documentclass[prl,twocolumn,showpacs,amsmath,amssymb]{revtex4}

\usepackage{graphicx}
\usepackage{enumerate}
\usepackage{times}
\usepackage{natbib}

\DeclareMathAlphabet{\bi}{OML}{cmm}{b}{it}

\long\def\symbolfootnote[#1]#2{\begingroup
\def\thefootnote{\fnsymbol{footnote}}\footnote[#1]{#2}\endgroup}

\begin{document}

\title{Statistical mechanics of ecosystem assembly}

\author{Jos\'e A.~Capit\'an$^1$,
Jos\'e A.~Cuesta$^1$ and Jordi Bascompte$^2$}

\affiliation{$^1${\it Grupo Interdisciplinar de Sistemas Complejos (GISC),
Departamento de Matem\'aticas, Universidad Carlos III de Madrid, E-28911
Legan\'es, Madrid, Spain}\\
$^2${Integrative Ecology Group, Estaci\'on Biol\'ogica de Do\~nana, Consejo
Superior de Investigaciones Cient\'{\i}ficas, c/ Americo Vespucio s/n,
E-41092 Sevilla, Spain}}

\begin{abstract}
We introduce a toy model of ecosystem assembly for which we are able to map out
all assembly pathways generated by external invasions.
The model allows to display the whole phase space in the form of an assembly
graph whose nodes are communities of species and whose directed links are transitions between
them induced by invasions. We characterize the process as a finite Markov chain
and prove that it exhibits a unique set of recurrent states (the endstate of
the process), which is therefore resistant to invasions. This also shows
that the endstate is independent on the assembly history. The model shares all
features with standard assembly models reported in the literature, with the
advantage that all observables can be computed in an exact manner.
\end{abstract}


\maketitle

Understanding why ecosystems are both stable and complex is still one of the
open questions in Ecology
\cite{mccann:2000}.
At present, it is widely accepted
that the dynamic process by which ecological communities are assembled is key to solve
this puzzle. It also has profound implications for conservation ---e.g.\ it can
shed light on how biodiversity may recover after major crisis, or
the influence it has on community stability.
Although ecosystem assembly has been studied experimentally
\cite{law:2000,warren:2003}, the bulk of studies are theoretical
\cite{post:1983,drake:1990,case:1990,law:1996,morton:1997,drossel:2001,loeuille:2005}.
These assembly models are but idealizations of the complex processes taking place
in real ecosystem assembly, which nevertheless implement the same mechanisms
acting in the formation of real ecosystems \cite{law:1999}. In this respect
these models are close in spirit to the general approach of statistical
physics of devising oversimplified paradigms which provide
insights into the real phenomena.

Previous work on ecosystem assembly made stochastic realizations of
sequential invasions based on a finite set of possible invaders
(known as ``regional species pool'' \cite{law:1996}) whose trophic
interactions are determined
in advance. These studies conclude: (i) at the end of the process, a final
endstate resistant to invasions is reached, which can be either a single
community or a set involving more than one
community between which the system fluctuates
\cite{morton:1997};
(ii) the average resistance of an ecosystem to be colonized increases in time,
and (iii) species richness also
increases in time. Then the assembly process favors increasing species richness
as well as stability, understood as resistance to invasions.

Successful as these models may be to provide insights into the formation of ecosystems,
we still lack a global picture of the process. The reason is at
least twofold. First, these model are defined as very complex stochastic processes
not amenable to analytical treatment, so that one can only hope to simulate
a limited set of realizations of the process and take averages on them.
This leaves some questions open such as, for instance, whether the endstate
depends on the assembly history. Although most evidence points to the uniqueness
of this endstate \cite{law:1999}, it is still a matter of discussion
\cite{fukami:2003}. Secondly, the size of the typical species pool
is always small, so that it has been argued 
\cite{case:1991}
that the exhaustion of good invaders
might justify the increasing resistance to invasion.

Our aim in this Letter is to propose a toy version of an assembly model which,
despite the stylized communities it deals with, still contains all the ingredients
of standard assembly models (something like the ``Ising model'' of ecosystem assembly)
and lacks some of its limitations (like the finiteness of the regional species
pool). Its major advantage over them is that
it allows to map out the whole ``phase space'' of the system, as well as to analyze
\emph{all} assembly pathways, something that permits
us to obtain exact results, to explore the
set of parameters characterizing different regimes, and to answer questions
out of reach of standard assembly models. The model
recovers all results mentioned above and
provides new ones ---the independence of the endstate from the assembly
history being the most prominent among them.

In what follows \emph{ecosystem} will refer to the system as
a whole, whereas \emph{community} will stand for any particular collection of
\emph{species,} i.e.\ any realization of the ecosystem.

Two are the ingredients that assembly models build upon:
A deterministic population dynamics for the species forming the community,
and a stochastic invasion process.
For the former we essentially adopt a mean-field-like
Lotka-Volterra model recently proposed to study coexistence in
competing communities and trophic level organization in food webs 
\cite{lassig:2001,bastolla:2005a,bastolla:2009} (these kind of models are called
\emph{neutral} in Ecology \cite{etienne:2007}, although neutral models
assume a stochastic population dynamics, unlike the one defined here). In this model, communities
are arranged in trophic levels, and species at level $\ell$ are assumed to
feed only on species at level $\ell-1$, and on all of them alike (mean field). Although omnivory
(i.e.\ feeding from lower levels) can be accounted for, it is still a matter of debate
how common it is \cite{bascompte:2005}, so we disregard it.
Accordingly, if $n_i^{\ell}$ denotes the population of species $i$ at level
$1\le \ell\le L$,
\begin{equation}
\begin{split}
\frac{\dot{n}_i^{\ell}}{n_i^{\ell}} &= -\alpha+\gamma_{+}N^{\ell-1}
- (1-\rho)n_i^{\ell} - \rho N^{\ell} -\gamma_{-}N^{\ell+1}, \\
\frac{\dot{n}_0}{n_0} &= R-n_0-\gamma_{-}N^1,
\end{split}
\label{eq:L-V}
\end{equation}
where $N^{\ell} \equiv \sum_{i=1}^{s_{\ell}} n_i^{\ell}$, $s_{\ell}$ being
the number of species at level $\ell$,
$\gamma_{+}$ controls the amount of energy available to reproduction
for each predation event, and $\gamma_{-}$ takes into account the mean damage
caused by predation to prey reproduction.
Following the standard rule of efficiency
on upwards energy transport to the next trophic level \cite{pimm:1982},
we assume that these two constants are related by $\gamma_+ = 0.1\gamma_-$.
Direct interspecific competition
is measured by $\rho < 1$, while intraspecific competition is set to one
to fix the population scale \cite{bastolla:2009}.
We regard all these species as consumers, and so they have a death
rate $\alpha>0$. For simplicity all parameters are assumed to be the same for
all species
(this assumption is harmless because the system has been shown to be structurally
stable against variations of the constants \cite{bastolla:2005a}).
The second of Eqs.~\eqref{eq:L-V} states that all species at the first
level (basal species) predate on a single resource whose ``amount'' is
given by $n_0$.
In the absence of basal species this resource reaches a steady value of
$n_0 = R$, thus this constant determines the maximum amount of resource available
to consumers. Finally, as real populations cannot be arbitrarily small, it
is required that extant species have a population $n_i^{\ell}>n_c$, the
\emph{extinction threshold}.
If a population falls below this value it is considered extinct and
removed from the community (see below).
Low populations are vulnerable against external variations or adverse
mutations \cite{pimm:1991}, and this stochastic effect is somehow accounted for
in the deterministic dynamics \eqref{eq:L-V} by the introduction of this viability condition.
Calculations have been carried out with $\alpha=1$, $n_c=1$, $\gamma_-=5$,
$\rho=0.3$ and $0<R\leq1700$. We have checked that variations of these
parameters do not change the qualitative picture.

Equations~\eqref{eq:L-V} have several equilibria, the main one being the
\emph{interior equilibrium.} In it all species of level $\ell$ have the same
population $n_{\ell}^*$. These are the solution to the system
\begin{equation}
\begin{split}
-\alpha =& \gamma_-N_{\ell+1}^*+\big[1+(s_{\ell}-1)\rho\big]N_{\ell}^*/s_{\ell}
-\gamma_+N_{\ell-1}^*, \\
R =& N_0^*+\gamma_-N_1^*,
\end{split}
\label{eq:interior}
\end{equation}
where $\ell=1,\dots,L$, $N_{\ell}^*=s_{\ell}n_{\ell}^*$, and we have the
constraints $s_0=1$, $s_{L+1}=0$.
The remaining equilibria are obtained by setting to zero any subset
of the populations and solving the linear system resulting from eliminating those
variables. Therefore one only needs the solutions of the linear systems
\eqref{eq:interior} for all possible choices of $s_{\ell}$ in order to
characterize the dynamical equilibrium of this model.
%
%
Interior equilibria are globally stable if, and only if,
$n_{\ell}^{*}>0$ for all $\ell=0,\dots,L$, because 
\begin{equation}
V\left(\left\{n_i^{\ell}\right\}\right) =
\sum_{\ell=0}^L \left(\frac{\gamma_{-}}{\gamma_{+}}\right)^{\ell}
\sum_{i=1}^{s_{\ell}} \left( n_i^{\ell} - n_{\ell}^{*}\log n_i^{\ell}\right)
\label{eq:lyapunov}
\end{equation}
is a Lyapunov function \cite{hofbauer:1998} for \eqref{eq:L-V}.
Therefore, given the set of species numbers $\{s_{\ell}\}_{\ell=1}^L$,
the corresponding community is viable and stable if, and only if,
$n_{\ell}^*>n_c$ for all $\ell=1,\dots,L$. Thus by solving \eqref{eq:interior}
for all choices of species numbers we can determine all viable
and stable communities. Let $\mathcal{F}$ denote the set of these communities.
Although the total size of a trophic level is not explicitly limited in any
way, the extinction threshold $n_c$ imposes a constraint and therefore
$\mathcal{F}$ is finite.

Equilibrium communities will undergo invasions which change their
composition. We shall assume that they take place at a uniform rate $\xi$.
Two hypotheses underlying this model, as well as most assembly models,
are: (i) the typical dynamical time is much smaller than $\xi^{-1}$, so
that communities are never invaded during transients, and (ii) the population
of invaders is small \cite{sax:2005}. Obviously, if the invasion rate
is too high \cite{fukami:2004,bastolla:2005b} or if invaders arrive with high
populations \cite{hewitt:2002}, the assembly dynamics ---and consequently the final
endstates--- will be seriously altered.
These situations are outside the scope of this model.

\begin{figure}
\begin{center}
\includegraphics[width=80mm,clip=true]{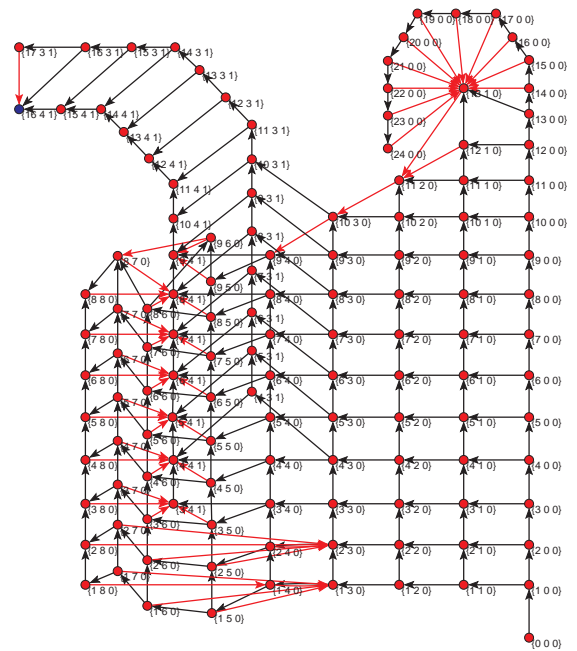}
\caption{(Color online.) Assembly graph obtained for $R=140$. It displays $130$
communities with up to three trophic levels. The numbers indicate how many
species are in each level. Black arrows represent accepted invasions; red arrows
represent transitions inducing a species loss. The only node in blue
(corresponding to the community $\{16,4,1\}$) represents
the endstate of the assembly process.%
}
\label{fig:assemblygraph}
\end{center}
\end{figure}

The invasion process goes as follows.
Consider a community $E\in\mathcal{F}$, with $L$ trophic levels, at its
equilibrium point. Invasions can occur at any level $\ell=1,\dots,L+1$, so we
randomly select $\ell$ and add a new species, with initial population $n_c$, at
level $\ell$ of the community $E$. Because of the global stability of our model, the extended
community evolves to the equilibrium given by \eqref{eq:interior} with $s_{\ell}$
replaced by $s_{\ell}+1$. If this equilibrium is viable, then we will have a new
community $E'$ consisting of $E$ plus the invader, and a transition will have
occurred from $E$ to $E'$. If, on the contrary, the population of the species of
at least one level falls below $n_c$, then extinctions will occur. It is 
unrealistic to extinguish the whole level, even though all its species have the
same population. In reality, if many species are threatened, by chance one of
them will be the first to become extinct. This fact may help the remaining
species to survive. Accordingly we shall extinguish species in an inviable level
as follows.
As we are monitoring the whole trayectory of the system, we can detect the
the moment when the first trophic level crossed $n_c$.
At that point we remove one species from that level and restart the evolution
from that point. We keep on removing one species at a time and restarting until the
new resulting equilibrium becomes viable. Two things can thus
happen: Either the first level to fall below $n_c$ is the invaded level, in which
case the invader is simply rejected and no transition occurs, or it is another level
that falls below $n_c$. In this case we end up with a new community $E''$, and a
transition will have occurred from $E$ to $E''$.

Starting from the empty community, $\varnothing$, and considering for every community
all possible invasions, we construct a directed graph, the \emph{assembly graph},
$\mathcal{G}$, whose nodes are elements of $\mathcal{F}$ and whose links are the
transitions obtained by the invasion process just described. Figure~\ref{fig:assemblygraph}
represents a typical assembly graph. From the viewpoint of statistical mechanics,
$\mathcal{G}$ is the phase space of our system.

\begin{figure}
\begin{center}
\includegraphics[width=80mm,clip=true]{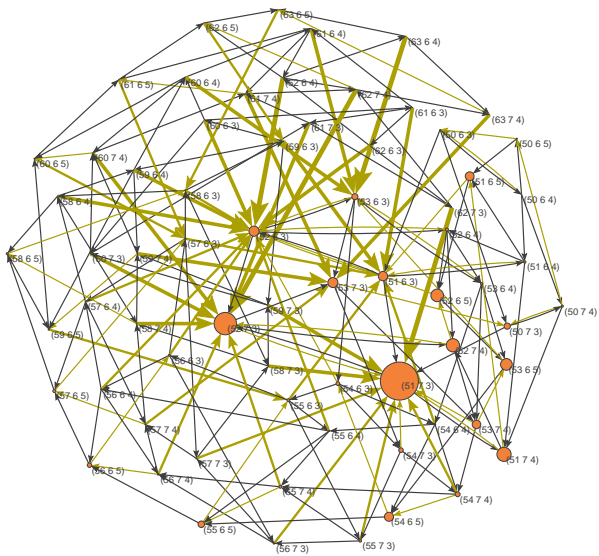}
\caption{(Color online.) Endstate of the assembly process obtained for $R=430$. It contains
$68$ communities (the assembly graph contains $3060$ communities). Black arrows represent
accepted invasions, whereas yellow ones represent transitions with species loss (width is
proportional to the number of extinctions). The diameter of the nodes is proportional to the
stationary probability of its community. Labels provide the number of species at each level.}
\label{fig:closedb}
\end{center}
\end{figure}

The assembly process becomes a Markov chain
if to every pair of communities $E$ and $E'$ we assign the transition probability
$P_{EE'} = \delta_{EE'} + \xi Q_{EE'}$, where $Q_{EE'}$ is the fraction of the $L+1$
different invasions of $E$ that lead to $E'$ ($Q_{EE'}\ne 0$ provided $\mathcal{G}$
contains the link $E\rightarrow E'$).
$P$ is the transition matrix of a
finite, aperiodic, Markov chain, so its states are either transient or
recurrent. There can be one or several subsets of recurrent states, the chain
being ergodic in each of them \cite{karlin:1975}. Every recurrent subset is a different
endstate of the assembly process. The assembly will depend on the history only if there
are at least two such recurrent subsets. Each recurrent subset has a stationary
probability distribution on its states.

It is worth noticing that the computation of $P$ is exact. This
means that we have a \emph{complete} and \emph{exact} description of the evolution of
the assembly process on the phase space. In particular this means that we can compute
the evolution of any observable in an exact ---albeit numeric--- manner, without resorting
to taking averages over realizations of the process. This is the most important difference
of this model with respect to all assembly models considered so far.

\begin{figure}
\begin{center}
\includegraphics[width=80mm,clip=true]{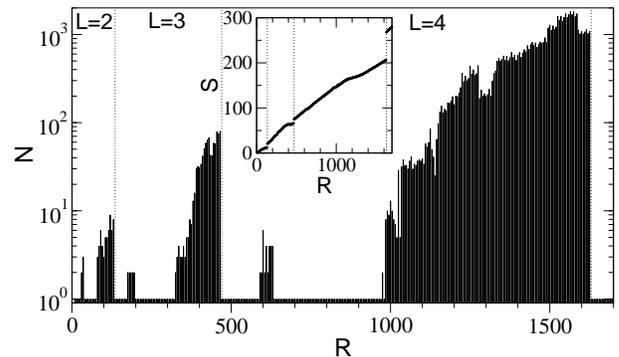}
\caption{Number of recurrent states of the Markov chain as a function of the
resource saturation $R$, obtained using the algorithm of Ref.~\cite{xie:1998}.
Inset: Mean number of species in the stationary state vs.\ $R$.}
\label{fig:recurrent}
\end{center}
\end{figure}

Let us now describe the results. First of all, the process has a unique set of recurrent states
for all $0<R\leq 1700$. This means that for this model we are able to prove that the endstate
does not depend on the assembly history \cite{drake:1990}. This result agrees with previous
evidence found in other assembly models \cite{morton:1997,law:1999} as well as in experiments
\cite{warren:2003}, where the same kind of assumptions about the invasion rate are made.
For some values of $R$ this endstate contains a single community
but typically it contains more than one, often very many.
Figure~\ref{fig:closedb} illustrates
one of these sets along with the transitions between its states.
At any time the ecosystem is realized by one of the communities of the set.
Invasions may induce transitions between communities of this set but
never lead outside it, hence the set as a whole is resistant to invasions.
The frequency with which a community is visited defines a stationary
probability on the set. Notice that only a few communities (rather similar to
each other) are visited with a high probability, so the net result is as
if the community were ``fluctuating''.
Figure~\ref{fig:recurrent} shows the number of communities
in the endstate 
vs.\ $R$.
One can see that as $R$ is increased towards
the onset of a new level appearance the number of communities increases considerably, to
drop down to just one once the new level has established.

Inset in Fig.~\ref{fig:recurrent} represents the mean number of species 
in the endstate
vs.\ $R$. The
dependence is basically linear, except for some dips near the onset of a new level
followed by a discontinuous jump once the new level is established. This behavior
reflects a top-predator effect:
Top predators control the populations of their preys, avoiding their extinction by
overconsumption of their resources
\cite{crooks:1999}. As a matter of fact we have checked that when a new level appears it
contains a single top predator and the number of species at lower levels rises.

\begin{figure}
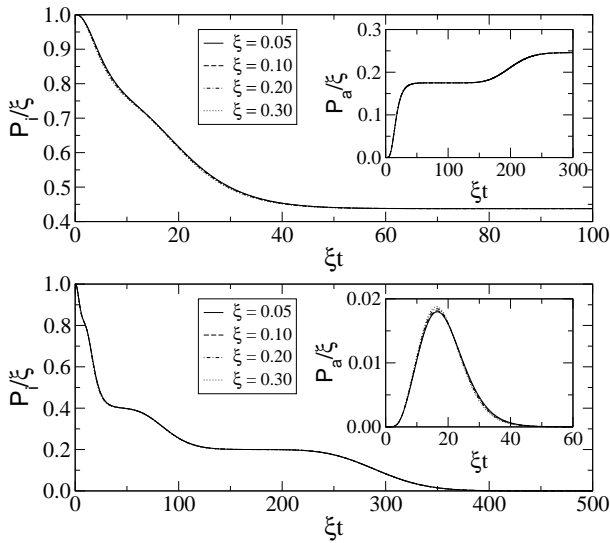

\begin{center}
\includegraphics[width=80mm,clip=true]{pinv-long.eps}
\includegraphics[width=80mm,clip=true]{pinv2-long.eps}
\caption{Probability of invasion, $P_i(t)$, vs.\ mean number of invasions, $\xi t$.
Inset: probability of reconfiguration after invasion, $P_a(t)$, vs.\ $\xi t$. Upper
pannel: $R=430$ (endstate of Fig.~\ref{fig:closedb}).
Lower pannel: $R=540$ (endstate with a single community).}
\label{fig:pinv}
\end{center}
\end{figure}

Other interesting observables are the probabilities of accepting the invader,
$P_i(t)$, and of undergoing a reconfiguration after the invasion, $P_a(t)$.
Both are obtained as $P_{i/a}(t) = \sum_{E}(\sum'_{E'}P_{EE'})(P^t)_{E\varnothing}$, 
where the inner sum runs over transitions from $E$ to $E'\ne E$ in which the
invader is accepted, or in which there is a reconfiguration of the
community, respectively. Figure~\ref{fig:pinv} illustrates them
for two typical cases: One with a complex endstate and another one
in which the endstate is a single community. Resistance to invasion
increases as the comminity assembles. 
Species richness can be computed similarly.
It grows monotonically up to saturation in the steady
state. Both features, increasing resistance to invasion and increasing species richness,
are common results of standard assembly models \cite{law:1996,morton:1997}.

In summary, this minimalistic assembly model exhibits the same behavior as
those reported in the literature, indicating that this behavior
is very robust and probably shared
also by real ecosystems.
Our model, however, provides a complete and exact description
of both, the set of microstates and the dynamical pathways
of the assembly process. So we are not limited, as in standard assembly
models, to compute averages over a small set of realizations of the process. 
To give a hint about what this means, we have calculated, for 
$R=300$ (a case with an endstate made of a single community of
three trophic levels and $50$ species), that there are
$\sim 10^{10}$ different
minimum-length pathways
leading from the empty to the endstate.
This number is nothing that a simulation can come close to.
This model allows us to prove rigorously that its endstate does not
depend on the assembly history. Whether this is a feature
that real ecosystems exhibit will, of course, depend on how well they fulfill
the assumptions about the invasion rate underlying this and other assembly models.
But a caveat should be made: As species of each trophic
level are indistinguishable, uniqueness (and hence
independence on history) refers only to the number of
species at each level.

We thank U.\ Bastolla, R.\ Law, S.\ C.\ Manrubia, A.\ Arenas, and J.\ Camacho 
for useful discussions. This work is funded by projects
MOSAICO (Ministerio de Educaci\'on y Ciencia) and MOSSNOHO-CM (Comunidad de Madrid),
by the European Heads of Research Councils, the
European Science Foundation, and the EC Sixth Framework Programme through an EURYI
award (J.\ B.), and by a contract from Comunidad de Madrid and Fondo Social Europeo
(J.\ A.\ Capit\'an).

\bibliography{ecology}

\end{document}